\documentclass[11pt]{article}
\usepackage{mathrsfs}
\usepackage{amsfonts}
\usepackage{amssymb}
\usepackage{bbm}
\usepackage{amsfonts,amssymb, mathrsfs, amsmath,   amssymb,  theorem,  float}
\usepackage{graphicx}  
\usepackage{color}
\usepackage{geometry}
\newtheorem{theorem}{Theorem}[section]

\newtheorem{cor}[theorem]{Corollary}
\newtheorem{lemma}[theorem]{Lemma}
\newtheorem{alg}[theorem]{Algorithm}

\newtheorem{defi}[theorem]{Definition}

\def\qed{\hfil {\vrule height5pt width2pt depth2pt}}

\def\qed{\hfil {\vrule height5pt width2pt depth2pt}}
\def\bref#1{(\ref{#1})}

\def\qed{\hfil {\vrule height5pt width2pt depth2pt}}

\def\C{\mathcal{C}}

\def\proof{{\noindent\em Proof.\,\,}}

\def\N{{\mathbb N}}
\def\Z{{\mathbb Z}}
\def\Q{{\mathbb Q}}

\def\C{{\mathbb C}}

\def\X{{\mathbb{X}}}

\def\F{{\mathbb{{F}}}}

\def\RB{{\mathcal{R}}}

\def\+{ \oplus}
\def\-{\ominus}
\def\*{\otimes}

\def\deg{\hbox{\rm{deg}}}

\def\modp{{\mathbf{mod}}}

\begin{document}

\title{Deterministic Interpolation of Sparse Black-box Multivariate Polynomials using Kronecker Type Substitutions\thanks{\quad Partially supported
by a grant from NSFC (11688101).}}
\author{Qiao-Long Huang$^{1,2}$ and Xiao-Shan Gao$^{1,2}$ \\
 $^{1}$KLMM, Academy of Mathematics and Systems Science\\
 Chinese Academy of Sciences, Beijing 100190, China\\
 $^{2}$University of Chinese Academy of Sciences, Beijing 100049, China}
\date{}

\maketitle

\begin{abstract}
\noindent
In this paper, we propose two new deterministic interpolation algorithms for a sparse multivariate polynomial given as a standard black-box  by introducing new Kronecker type substitutions.
Let $f\in \RB[x_1,\dots,x_n]$ be a sparse black-box polynomial with a degree bound $D$.
%
When $\RB=\C$ or a finite field, our algorithms either have better bit complexity
or better bit complexity in $D$ than existing deterministic algorithms.
In particular, in the case of deterministic algorithms for standard black-box models, our second algorithm has the current best complexity in $D$ which is the dominant factor in the complexity.

\vskip10pt
\noindent{\bf Keywords}.
Sparse polynomial interpolation,
Kronecker substitution,
black-box model.
\end{abstract}

\section{Introduction}
Sparse polynomial interpolation is a basic computational problem
which attracts lots of attention recently.
There exist two main classes of sparse polynomial interpolation algorithms: the {\em direct approaches}  \cite{3,32,14,15,3,32,kal-e1},  which recover the exponents of the multivariate polynomials  directly
and the {\em reduction approaches} \cite{1,5,9,8, huang-slp,13},
which reduce multivariate interpolations to univariate interpolations.

The sparse polynomial can be given as a standard black-box model
or more special models such as the straight-line program (SLP) model.
Since the value of a polynomial of degree $D$ at any point other than $0,\pm 1$ will have $D$ bits or more, any algorithm whose complexity is polynomial in $\log D$ cannot perform such an evaluation over $\Q$ or $\Z$. Even for polynomials over the  finite field $\F_q$, there exist no interpolation algorithms whose complexity is polynomial in $\log D$ and $\log q$ for the standard black-box model.
For special models such as the SLP  model \cite{2,21,s10,5,8,huang-slp}, the  precision accuracy black-box model \cite{AM95,lee-new,M95,14},
and the modular black-box model \cite{GR10,BJ14}, there exist
interpolation algorithms whose complexity is polynomial in $\log D$.

In this paper, we propose two new deterministic interpolation algorithms for
standard black-box multivariate polynomials by introducing new Kronecker type substitutions.

\subsection{Main results}
Let $f\in \RB[x_1,\dots,x_n]$ be a sparse polynomial given by a black-box $B_f$,
where $\RB$ is an integral domain.
Suppose that $f$ has a degree bound $D$ and a term bound $T$.
Our algorithms are based on the following Koronecker type substitutions:
\begin{eqnarray}
f^{\modp}_{(d,p)}   &=&  f(x,x^{d},\dots,x^{d^{n-1}})\ \modp \ (x^p-1)\label{sb-3}\\
f_{(d,p)}  &=&  f(x,x^{\modp(d,p)},\dots,x^{\modp(d^{n-1},p)})\label{sb-1}\\
f_{(d,p,k)}  &=&
f(x,\dots,x^{\modp(d^{k-1},p)+p},\dots,x^{\modp(d^{n-1},p)})\label{sb-2}
\end{eqnarray}
where $d,p\in\Z_{>0}$.
%

In our first algorithm, we first interpolate the univariate polynomials $f_{(d,p)}$ for a fixed prime $p$ and $d=1,\dots,4(n-1)(T-1)+1$, and then recover $f$ from these $f_{(d,p)}$.
In this case, $f$ is recovered with the {\em base changing Kronecker substitutions}: $x_i = x^{\modp(d^{i-1},p)}, d=1,\dots,4(n-1)(T-1)+1$.

In our second algorithm, we first interpolate the univariate polynomials
$f_{(D,p_k)},k=1,\dots,N$ and then recover $f$ from these $f_{(d,p)}$,
where $N$ is an integer of size $O(nT\log D)$
and $p_k$ are different prime numbers.
In this case, $f$ is recovered  with the {\em modulus changing Kronecker substitutions}: $x_i = x^{\modp(D^{i-1},p_k)},k=1,\dots,N$.
%

 \begin{table}[htb]\footnotesize
\centering
 \scalebox{0.90}[0.90]{%
\begin{tabular}{c|c|c|c|c}
 &Probes &Arithmetic &Height of  &Bit  \\
 &       &operations& data &complexity \\\cline{1-5}
Ben-or $\&$ Tiwarri \cite{3,32}
&$T$&$nT^2D$&$TD$&$nT^3D^2$\\
Zippel \cite{13}&$nT^2D$ &$nT^3D$&$TD$&$nT^4D^2$\\
Klivans $\&$ Spielman \cite{9}
&$nT^3$&$nT^3\log^2 D$ & $TD\max(nT^2,D)$ & $nT^4D\max(nT^2,D)$\\
This paper (Cor. \ref{cor-b0-1})&$nT^2$&$nT^2\log^3D$&$TD\max(nT,D)$&$nT^3D\max(nT,D)$\\
This paper (Cor. \ref{cor-b0-2})&$nT^2\log D$&$nT^2\log^3 D$&$nT^2D$&$n^2T^4D$\\
%
\end{tabular}}
\caption{A ``soft-Oh" comparison of deterministic interpolation algorithms over $\C$}\label{tab-1}
\end{table}

Table \ref{tab-1} is a comparison with other deterministic algorithms when $\RB=\C$,
where ``Probes" is the number of calls to $B_f$.
We assume that the size of the coefficients is $O(1)$ to simplify the results.
From the table, our second algorithm has the lowest complexity in $D$.
Probabilistic algorithms and algorithms for special black-box models,
such as the SLP model, are not compared here.

\begin{table}[ht]\footnotesize
\centering
 \scalebox{0.9}[0.9]{%
\begin{tabular}{c|c|c|c}
&Probes&Bit complexity&Size of $\F_q$ \\\cline{1-4}
Grigoriev-Karpinski-Singer \cite{101}&&$n^2T^6\log^2(q)+q^{2.5}\log^2q$& $q\geq O^\thicksim(n^2T^2)$\\
 Klivans $\&$ Spielman \cite{9}&$nT^3$&$nT^2D\max(nT^2,D)\log q$&$ q\geq O^\thicksim(D\max(nT^2,D))$\\
This paper (Cor. \ref{cor-bq-1}) &$nT^2$&$nTD\max(nT,D)\log q$&$q\geq O^\thicksim(D\max(nT,D))$\\%
This paper (Cor. \ref{cor-bq-2}) &$nT^2\log D$&$n^2T^2D\log q$&$q\geq O^\thicksim(nDT)$
\end{tabular}}
\caption{``Soft-Oh"  comparison of deterministic interpolation algorithms $\mathbf{F}_q$ }\label{tab-2}
\end{table}

Table \ref{tab-2} is a comparison with other deterministic algorithms over finite fields,
where ``Size of $\F_q$'' means that this algorithm works for $\F_q$ satisfying the condition.
If the elements of the extension field of $\F_q$ can be used, algorithms work for any finite field.
From the table, our first algorithm has better complexity than that in \cite{9}.
%
In \cite{101}, it is assumed that $\deg(f,x_i)< q-1$.
Under the same assumption, $D\le n (q-1)$ and our second algorithm has complexity
$O^\thicksim(n^3T^2q)$.
So, our second algorithm is linear in $q$ and the algorithm in \cite{101}
is linear in $q^{2.5}$. Also, our algorithm has worse complexity in $n$
but better complexity in $T$.
There are recent works on the probabilistic algorithms over  finite fields \cite{1,16,huang-bot,17,13}, which have better complexities than deterministic algorithms.

Note that the size of $f$ is about $O(nT\log D)$.
Therefore, $D$ is the exponential part in the complexity and
hence the dominant factor.
Our second algorithm has the best complexity in $D$
among the deterministic algorithms for standard
black-box multivariate polynomials.

\subsection{Technique contribution and relation with existing work}

Our algorithm has three major steps.
First, we compute a set of univariate polynomials $f^{\modp}_{(d,p)}$ and show that if $f^{\modp}_{(d_0,p_0)}$ has the maximal number of terms, then at least half of the
terms of $f$ do not merge or collide with other terms of $f$ in $f^{\modp}_{(d_0,p_0)}$.
The
$d_0$ in the base changing Kronecker substitution and
$p_0$ in the modulus changing Kronecker substitution
are called ``ok" base and ``ok" prime, respectively.
Second, we interpolate extra $n$ univariate polynomials $f_{(d_0,p_0,k)}$
to find a set of terms containing these non-colliding terms.
Finally, we give a criterion to test whether $f$ contains a given term
and use this criterion to pickup the terms of $f$ from the terms found in the second step.

The main contributions of this paper are: new Kronecker substitutions,
criteria for term testing, and methods to find ``ok" base and ``ok" prime.

In the rest of this section, we compare our algorithms with \cite{5,8,2,9,huang-slp}.
Our work builds on and is inspired by these works.

The idea of base changing Kronecker substitutions given in Section 3
is introduced in this paper for the first time.
The idea of modulus changing Kronecker substitutions used in Section 4 was given in \cite{5,9,huang-slp}, but the method in this paper is different from them and will be explained below.

The substitution $f^{\modp}_{(D,p_j)}$ in \bref{sb-3} was introduced by
Grag and Schost \cite{5} to interpolate  an SLP polynomial $f$ by recovering $f$ from
$f^{\modp}_{(D,p_j)}$ for $O(T^2\log D)$ different primes $p_j$.
Our second method is similar to this method, but works for black-box models
and has the following differences.
First, for a black-box polynomial $f$, $f^{\modp}_{(D,p)}$ cannot be computed
to keep all degrees in $x$ less than $p$.
In order to work for black-box polynomials, we use the substitution $f_{(D,p)}$ to compute $f^{\modp}_{(D,p)}= f_{(D,p)}\ \modp\ (x^p-1)$.
Second, we interpolate $f^{\modp}_{(D,p_j)}$ for $O(T\log D)$ different primes $p_j$.
Third, we give a new criterion to check whether a term belongs to $f$.
Finally, a new Kronecker substitution $f_{(d,p,k)}$ is introduced to recover the exponents.

The substitution $f_{(d,p)}$  in \bref{sb-1} was introduced by Klivans-Spielman \cite{9} to interpolate black-box polynomials.
Instead of $f_{(d,p,k)}$, they used
the substitution $f_{(KS)}=f(q_1x,$ $q_2x^{\modp(d,p)},\dots,$ $q_nx^{\modp(d^{n-1},p)})$
for different primes $q_i$.
Our substitution $f_{(d,p,k)}$ has the following advantages:
(1) For the complex field, the size of coefficients is not changed after our substitution, while the size of coefficients in $f_{(KS)}$ is increased by a factor of $D$.
(2) Our algorithm works for general rings, while
the substitution in \cite{9} needs an element $\omega\in \RB$ such that $\omega^i\neq 1,i=1,2,\dots,D$.

The substitution $f_{(d,p,k)}$  in \bref{sb-2} was introduced in \cite{huang-slp}
for interpolating SLP polynomials. The method based on
modulus changing Kronecker substitutions given in Section 4 of this paper
could be considered as a generalization of the method given in  \cite{huang-slp}
from SLP model to black-box model.
The major difference is that for black-box model,
$f^{\modp}_{(D,p)}$ is computed from $f_{(D,p)}\ \modp\ (x^p-1)$,
where the degrees of the intermediate polynomials
were bounded by $pD$ instead of $p$.
As a consequence, we need to give new  methods for the two key ingredients
of the algorithm: the criterion for term testing and
the method of finding the ``ok" prime.

In Arnold, Giesbrecht, and Roche \cite{21}, the concept of
``ok" prime is introduced.
A prime $p$ is ``ok"  if at most $\frac38T$ terms of $f$ are collisions in $f^{\modp}_{(D,p)}$.  The ``ok" prime in \cite{21} is probabilistic.
In this paper, we give similar notions of ``ok" base and ``ok" prime,
and determinist methods to compute them.
Also, the randomized Kronecker substitution is used in \cite{2},
while our Kronecker reduction is deterministic.
%

For interpolation over finite fields,
Grigoriev-Karpinski-Singer \cite{101} gave the first deterministic algorithm in finite fields. Prony's algorithm is dominated by the cost of discrete logarithms in $\F_q$.
%
%
Ben-or $\&$ Tiwarri's algorithm and Zippel's algorithm
meet the {\em Zero Avoidance Problem} \cite{16,13,17}.
Klivans-Spielman's algorithm needs to compute a factorization and thus only works for the integral domain which contains an element $\omega$ such that $\omega^i\neq 1,i=1,2,\dots,D$.
Our algorithms are more general comparing to existence methods in the sense that they do not need to solve the zero avoidance problem, the discrete logarithm problem, and the factorization problem, which is possible mainly due to the
substitutions $f_{(d,p)}$ and $f_{(d,p,k)}$.
%

\section{Three Koronecker type substitutions}
Throughout this paper, let
$f=c_1m_1+\cdots+c_t m_t\in \RB[\X]$ be a black-box multivariate polynomial with terms $c_im_i$,
where $\RB$ is an integral domain, $\X = \{x_1,\ldots,x_n\}$ be a set of $n$ indeterminates,
and $m_i,i=1,\dots,t$ are distinct monomials.
Denote $\#f=t$ to be the number of terms of $f$,
$\deg f=t$ to be the total degree of $f$,
and  $M_f=\{c_1m_1,\dots,c_tm_t\}$ to be the set of terms of $f$. Let  $D,T\in\N$ such that $D> \deg(f)$ and $ T\ge\#f$.

Let $d\in\N,p\in\N_{>0}$ and $x$ a new indeterminate.
Consider the univariate polynomials in $\RB[x]$:
\begin{eqnarray}
f_{(d,p)} \!\!\!&=&\!\!\! f(x,x^{\modp(d,p)},\dots,x^{\modp(d^{n-1},p)})\label{eq-f11}\\
f_{(d,p,k)} \!\!\!&=&\!\!\! f(x,\dots,x^{\modp(d^{k-1},p)+p},\dots,x^{\modp(d^{n-1},p)})\label{eq-f13}\\
f^{\modp}_{(d,p)} \!\!\!&=&\!\!\! f_{(d,p)}\ \modp \ (x^p-1) = f_{(d,p,k)}\ \modp \ (x^p-1)\label{eq-f12}
\end{eqnarray}
where $k=1,2,\dots,n$. $\bref{eq-f11}$ comes
from the modified Kronecker substitutions
$x_i=x^{\modp(d^{i-1},p)}$, $i=1,2,\dots,n$ and $\bref{eq-f13}$ comes from another modified Kronecker substitutions $x_i=x^{\modp(d^{i-1},p)}$,$i=1,2,\dots,n,i\neq k, x_k=x^{\modp(d^{k-1},p)+p}$,
respectively.
We have the following key concept.
\begin{defi}\label{def-1}
A term $cm\in M_f$ is called a collision in $f^{\modp}_{(d,p)}$ ($f_{(d,p)}$ or $f_{(d,p,k)}$) if  there exists
an $aw\in M_f\backslash \{cm\}$ such that $m^{\modp}_{(d,p)}\!=\!w^{\modp}_{(d,p)}$
($m_{(d,p)}\!=\!w_{(d,p)}$ or $m_{(d,p,k)}\!=\!w_{(d,p,k)}$).
%
\end{defi}

The following fact is obvious.

\begin{lemma}\label{lm-2}
 Let $cm\in M_f$. If $cm$ is not a collision in $f_{(d,p)}^{\modp}$, then  $cm$ is not a collision in $f_{(d,p)}$, and  $cm$ is not a collision in any $f_{(d,p,k)}$ for $k=1,2,\dots,n$.
\end{lemma}

Let
\begin{equation}\label{eq-1}
 f_{(d,p)}^{\modp}=a_1x^{d_1}+a_2x^{d_2}+\cdots+a_r x^{d_r}\quad (d_1<\cdots<d_r)
\end{equation}
Since $f_{(d,p)}\ \modp\ (x^p-1)=f_{(d,p,k)}\ \modp\ (x^p-1)=f^{\modp}_{(d,p)}$, for $k=1,2,\dots,n$, we can write
  \begin{eqnarray}
f_{(d,p)}&=&f_1+f_2+\cdots+f_r+g\label{eq-k11}\\
f_{(d,p,k)}&=&f_{k,1}+f_{k,2}+\cdots+f_{k,r}+g_k\label{eq-k12}
\end{eqnarray}
where $f_i\ \modp\ (x^p-1)=f_{k,i}\ \modp\ (x^p-1)=a_ix^{d_i}$, $g\ \modp\ (x^p-1)=g_k\ \modp\ (x^p-1)=0$.
We define the following key notation.
\begin{eqnarray}
&&\hbox{TS}^f_{d,p,D} =\{a_i x_1^{e_{i,1}}\cdots x_n^{e_{i,n}}| a_i \hbox{ is from } \bref{eq-1}, \hbox{ and}  \cr
 &&\quad\quad\quad\hbox{T}_1:
 \hbox{In } \bref{eq-k11} \hbox{ and }  \bref{eq-k12}, f_i=a_ix^{\gamma_i},f_{k,i}=a_ix^{\beta_{k,i}},k=1,2,\dots,n.\label{eq-uf2}\\ 
 &&\quad\quad\quad \hbox{T}_2: e_{i,k}=\frac{\beta_{k,i}-\gamma_i}{p}\in\N,k=1,2,\dots,n.\cr
 &&\quad\quad\quad \hbox{T}_3:  \gamma_i=\sum_{k=1}^n e_{i,k}\,\modp(d^{k-1},p) \}.\nonumber\\
 &&\quad\quad\quad \hbox{T}_4:  e_{i,1}+\cdots+ e_{i,n}< D\}.\nonumber
\end{eqnarray}

Similar to \cite[Lemma 5.3]{huang-slp}, we can prove the following result about $\hbox{TS}^f_{d,p,D}$.

\begin{lemma}\label{lm-7}
Let $f=\sum_{i=1}^tc_im_i\in\RB[\X]$, $d,p\in\Z_{>0}$, $D>\deg (f)$. If $c_im_i$ is not a collision in $f^{\modp}_{(d,p)}$, then $c_im_i\in \hbox{TS}^f_{d,p,D}$.
\end{lemma}

The following algorithm is used to find the set $\hbox{TS}^f_{d,p}$.
\begin{alg}[TSTerms]\label{alg-1}
\end{alg}

{\noindent\bf Input:}
Univariate polynomials $f^{\modp}_{(d,p)},f_{(d,p)},f_{(d,p,k)}$ ($k=1,\dots,n$) in $\RB[x]$, a positive integer $d$, a prime $p$,  $D> \deg (f)$.

{\noindent\bf Output:} $\hbox{TS}^f_{d,p,D}$.

\begin{description}
\item[Step 1:]
Write $f_{(d,p)}^{\modp},f_{(d,p)},f_{(d,p,k)}$ in the following form
\begin{eqnarray*}
f_{(d,p)}^{\modp} &=&a_1x^{d_1}+\cdots+a_rx^{d_r}\\
f_{(d,p)}&=&a_1x^{\gamma_1}+\cdots+a_{\alpha}x^{\gamma_{\alpha}}+g\\
f_{(d,p,k)}&=&a_1x^{\beta_{k,1}}+\cdots+a_{\alpha}x^{\beta_{k,{\alpha}}}+g_k
\end{eqnarray*}
such that for $i=1,\dots,\alpha,k=1,\dots,n$,  $\modp(\gamma_i,p)=\modp(\beta_{k,i},p)=d_i$ and $g^{\modp}_{(d,p)} = (g_k)^{\modp}_{(d,p)} =0$.
%

\item[Step 2:] Let $S=\{\}$.
For $i=1,2,\dots,\alpha$, do
    \begin{description}
    \item[a:]
    for $k=1,2,\dots,n$, do

 let $e_{i,k}=\frac{\beta_{k,i}-\gamma_i}{p}$.
 If $e_{i,k}$ is not in $\N$, then break;
    \item[b:] If $\sum_{k=1}^ne_{i,k}\modp(d^{k-1},p)\neq \gamma_i$, then break;

    \item[c:] If $\sum_{k=1}^ne_{i,k}\geq D$, then break;

    \item[d:]  Let $S=S\bigcup\{a_ix_1^{e_{i,1}}\cdots x_n^{e_{i,n}}\}$.
    \end{description}

\item[Step 3:] Return $S$.
\end{description}

Similar to  \cite[Lemma 5.5]{huang-slp}, we can prove the following lemma.
\begin{lemma}\label{the-3}
Algorithm \ref{alg-1} needs $O(nT)$ arithmetic operations in $\RB$ and $O^\thicksim(nT\log (pD))$ bit operations.
\end{lemma}
%

\section{An interpolation algorithm based on base changing Kronecker substitution}
In this section, we give the first interpolation algorithm.
The basic idea of the algorithm is to recover $f$ from
$f_{(d,p)}$ for a fixed prime $p$ and  sufficiently many values for $d$,
that is, we will change the  degree base $d$ in the Kronecker substitution $x^{d^i}$.

\subsection{A criterion for term testing}
In this subsection, we give a criterion to check whether
a given term belongs to a polynomial.
Let $\F_p$ be the finite field with $p$ elements.
%


\begin{lemma}\label{lm-4}
Let $L_i=a_{i,1}+\cdots+a_{i,n}x^{n-1}\in \F_p[x]\setminus\{0\},i=1,\dots,l$,  $p$ a prime such that $p\geq\max\{n, (n-1)l\}$. If $\delta$ is an integer satisfying $(n-1)l\le \delta\le p$, then there exist at least $\delta-(n-1)l$ integers $k$ in $[1,\delta]$ such that $L_i(k)\neq 0$, for all $i=1,2,\dots,l$.
\end{lemma}
\proof
For each $L_i$, since $\deg(L_i)=n-1$, there exist at most $(n-1)$ integers $k$ in $[1,\delta]$ such that  $L_i(k)=0$. Since we have $l$ nonzero functions $L_i$, there exist at most $(n-1)l$ integers $k$ in $[1,\delta]$ such that for some $L_i(k)= 0$. So the rest integers in $[1,p]$ does not vanish $L_i,i=1,2,\dots,l$.\qed

\begin{lemma}\label{lm-5}
Let $f=\sum_{i=1}^tc_im_i\in\RB[\X]$,
$T\geq \#f,$ $D>\deg(f)$, $\delta_1=(n-1)(T-1)$,
and  $p$ a prime such that  $p\geq \max\{n,\delta_1,D\}$.
Then for any integer $\delta$ satisfying $\delta_1 \leq \delta\leq p$
and any integer $i\in[1,t]$,  there exist at least $\delta-\delta_1$ integers $d$ in $[1,\delta]$,
such that $c_im_i$ is not a collision in $f^{\modp}_{(d,p)}$.
\end{lemma}

\proof
If $t=1$, the proof is obvious. Now we consider the case $t\geq 2$.
It suffices to consider   $c_1m_1$.
Assume $m_i=x_1^{e_{i,1}}x_2^{e_{i,2}}\cdots x_n^{e_{i,n}},i=1,2,\dots,t$.
Let
$L_s(x)=(e_{1,1}-e_{s,1})+\cdots+(e_{1,n}-e_{s,n})x^{n-1}\ \modp\ \ p$,
where $s=2,3,\dots,t$.
Since $m_1,m_s$ are different monomials, at least one of $e_{1,k}-e_{s,k}\neq 0,k\in\{1,2,\dots,n\}$.  Since $p\geq D$ and $|e_{1,i}-e_{s,i}|<D$, we have $L_s(x)\neq 0$.
We claim that if $d\in[1,\delta]$ such that $L_s(d)\neq 0$, then $m^{\modp}_{1(d,p)}\neq m^{\modp}_{s(d,p)}$. Since $\deg (m^{\modp}_{1(d,p)})=\sum_{k=1}^ne_{1,k}d^{k-1}\ \modp\ p$, $\deg (m^{\modp}_{s(d,p)})=\sum_{k=1}^ne_{s,k}d^{k-1}\ \modp\ p$, we have $m^{\modp}_{1(d,p)}\neq m^{\modp}_{s(d,p)}$. We proved the claim.
By Lemma \ref{lm-4}, there are $\geq\delta-(n-1)(T-1)\geq \delta-\delta_1$ integers in $[1,\delta]$ such that all $L_s(x),s=2,3,\dots,t$ are non-zero.
\qed

Now we give a criterion for testing whether a term $cm$ is in $M_f$.

\begin{theorem}\label{the-1}
Let $f=\sum_{i=1}^t c_im_i\in\RB[\X]$, $T\geq \# f,D>\deg(f)$, $\delta_1=(n-1)(T-1)$, $\delta_2=(n-1)T$,
and $p$ a prime with $p\geq \max\{n,\delta_1+\delta_2+1,D\}$.
For a term $cm$ satisfying $\deg (m)<D$, $cm\in M_f$ if and only if there exist at least $\delta_2+1$
integers $d\in[1,\delta_1+\delta_2+1]$ such that $\#(f-cm)^{\modp}_{(d,p)}<\#f^{\modp}_{(d,p)}$.
%
\end{theorem}
\proof
Let $cm\in M_f$. If $d$ is an integer such that $cm$ is not a collision in $f^{\modp}_{(d,p)}$, then   $\#(f-cm)^{\modp}_{(d,p)}=\#f^{\modp}_{(d,p)}-1$, and hence $\#(f-cm)^{\modp}_{(d,p)}<\#f^{\modp}_{(d,p)}$.
By Lemma \ref{lm-5}, there exist at most $\delta_1+\delta_2+1-\delta_1=\delta_2+1$ integers $d\in[1,\delta_1+\delta_2+1]$
such that $cm$ is not a collision in $f^{\modp}_{(d,p)}$. So there exist at least $\delta_2+1$ integers $d$ such that $\#(f-cm)^{\modp}_{(d,p)}<\#f^{\modp}_{(d,p)}$.

For the other direction, assume $cm\notin M_f$. We show there
exist at most $\delta_2$ integers $d\in[1,\delta_1+\delta_2+1]$ such that $\#(f-cm)^{\modp}_{(d,p)}<\#f^{\modp}_{(d,p)}$.
Consider two cases:
Case 1: $m$ is not a monomial in $f$.
Then $f-cm$ has at most $T+1$ terms.
By Lemma \ref{lm-5}, there exist  at least $\delta_1+\delta_2+1-\delta_2=\delta_1+1$ integers $d\in[1,\delta_1+\delta_2+1]$
such that $cm$ is not collision in $(f-cm)^{\modp}_{(d,p)}$.
So there are at least $\delta_1+1$ integers $d$ such that $\#(f-cm)^{\modp}_{(d,p)}>\#f^{\modp}_{(d,p)}$.
So there are at most $\delta_1+\delta_2+1-(\delta_1+1)=\delta_2$ integers $d$ such that $\#(f-cm)^{\modp}_{(d,p)}<\#f^{\modp}_{(d,p)}$.
Case 2: $m$ is a monomial in $f$, but $cm$ is not a term in $f$. Then $f-cm$ has the same number of terms as $f$.
Let the term of $f$ with monomial $m$ be $c_1m$.
By Lemma \ref{lm-5}, there exist at least $\delta_1+\delta_2+1-\delta_1=\delta_2+1$ integers
$j$ such that $(c_1-c)m$ is not collision in $(f-cm)^{\modp}_{(d,p)}$.
So there are at least $\delta_2+1$ integers $j$ such that $\#(f-cm)^{\modp}_{(d,p)}=\#f^{\modp}_{(d,p)}$.
Since $\delta_1+\delta_2+1-(\delta_2+1)=\delta_1<\delta_2$, there are at most $\delta_2$ integers $d$ such that $\#(f-cm)^{\modp}_{(d,p)}<\#f^{\modp}_{(d,p)}$.
\qed

%
%

\subsection{Find an ``ok'' degree}
In this subsection, we show how to find an ``ok" degree $d$ such that half of terms in $f$ do not collide in $f^{\modp}_{(d,p)}$ for certain prime $p$.

\begin{lemma}\label{lm-51}\cite{huang-bot}
Let $B_j,j=1,2,\dots,s$ be nonempty sets of integers  and $a_i,i=1,2,\dots,t$ all the distinct elements in $\cup_{j=1}^s B_j$. Let $c$ be the number of $a_i$ satisfying $a_i\in B_j$ and $\#B_j\geq 2$ for some $j$.
Then $t-c\le s$ and for $s_1\in[t-c,s]\cap\N$, we have $(t-s_1)\leq c\leq2(t-s_1)$.
%
\end{lemma}

Denote $\mathcal{C}^f_{(d,p)}$ to be the number of collision terms of $f$ in $f^{\modp}_{(d,p)}$.
Then, we have
\begin{lemma}\label{cor-2}
Let $f\in\RB[\X]$,  $\#f^{\modp}_{(d_1,p)}=s_0$, and
$\#f^{\modp}_{(d_2,p)}=s$. If $s_0\geq s$, then $\mathcal{C}^f_{(d_1,p)}\leq 2\mathcal{C}^f_{(d_2,p)}$.
\end{lemma}
\proof
Assume $f^{\modp}_{(d_1,p)}=a_1x^{e_1}+\cdots+a_{s_0}x^{e_{s_0}},e_i\neq e_j$, when $i\neq j$. Let $f=f_1+\cdots+f_{s_0}+g$, where $f^{\modp}_{i(d_1,p)}=a_ix^{e_i},i=1,\dots,s_0$ and $g^{\modp}_{(d_1,p)}=0$. Let $B_i,i=1,\dots,s_0$ be the sets of terms in $f_i$ and $B_0$ be the set of terms in $g$. So by Lemma \ref{lm-51}, we have $(t-s_0)< \mathcal{C}^f_{(d_0,p)}\leq2(t-s_0)$. By the same reason, we have $(t-s)\leq\mathcal{C}^f_{(d_2,p)}\leq2(t-s)$.
Now $\mathcal{C}^{f}_{(d_1,p)}\leq 2(t-s_0)\leq2(t-s)\leq 2\mathcal{C}^{f}_{(d_2,p)}$. The lemma  is proved.\qed

\begin{lemma}\label{lm-6}
Let $f=\sum_{i=1}^tc_im_i\in\RB[\X]$, $m_i=x_1^{e_{i,1}}\cdots x_n^{e_{i,n}}$, $D>\deg(f)$, $p$ a prime such that $p\geq \max\{n,D\}$, $A_{u,v}=\sum_{s=1}^n(e_{u,s}-e_{v,s})x^{s-1}\ \modp\ p$, where $u,v\in\{1,\dots,t\}$, and $u<v$.
If $\mathcal{C}^f_{(d_0,p)}=s$ and $d_0$ is an integer in $[1,p]$,
then  at least $\lceil\frac s2\rceil$ of $A_{u,v}$ satisfy $A_{u,v}(d_0)=0$.
%
\end{lemma}
\proof
 Let $n_i$ be the number of collision blocks with $i$ terms in $f^{\modp}_{(d_0,p)}$. Assume $c_{j_1}m_{j_1}+\cdots+c_{j_i}m_{j_i}$ is a collision block with $i$ terms. For any $u,v\in\{j_1,\dots,j_i\},u<v$, we have $m^{\modp}_{u(d_0,p)}= m^{\modp}_{v(d_0,p)}$.
So $(e_{u,1}+\cdots+e_{u,n}{d_0}^{n-1})\ \modp\ p=(e_{v,1}+\cdots+e_{v,n}{d_0}^{n-1})\ \modp\ p$, which implies that $d_0$ is a root of $A_{u,v}$.
There exist $C_i^2=\frac{i(i-1)}{2}$  such pairs $(u,v)$ and $n_i$ such collision blocks. Let $K=\sum_{i=1}^t\frac{1}{2}(i^2-i)n_i$. So there are $K$ different $A_{u,v}$ with root $d_0$.
Now we give a lower bound of $K$. First we see that $t=\sum_{i=1}^tin_i,s=\sum_{i=2}^tin_i$.
$K=\sum_{i=1}^t\frac{1}{2}(i^2-i)n_i=\frac{1}{2}\sum_{i=1}^ti^2n_i-\frac{1}{2}\sum_{i=1}^tin_i=\frac{1}{2}\sum_{i=1}^ti^2n_i-\frac{1}{2}t=\frac12n_1+\frac{1}{2}\sum_{i=2}^ti^2n_i-\frac{1}{2}t\geq
\frac12n_1+t-n_1-\frac12t=\frac12 t-\frac12n_1=\frac12 s$.
 Since $K$ is an integer, $K\geq \lceil\frac12 s\rceil$. \qed

\begin{theorem}\label{the-2}
Let $f=\sum_{i=1}^tc_im_i\in\RB[\X]$, $T\geq \# f,D>\deg f$, $\delta_1=(n-1)(T-1)$, and $p$ be a prime such that $p\geq \max\{n,4\delta_1+1,D\}$. Let $d_0$ be an integer in $[1,4\delta_1+1]$ such that $\# f^{\modp}_{(d_0,p)}\geq \# f^{\modp}_{(d,p)}$ for all $d=1,\ldots,4\delta_1+1$. Then  at least $\lceil\frac t 2\rceil$ terms of $f$ do  not a collide in $f^{\modp}_{(d_0,p)}$.
\end{theorem}
\proof
If $t=1$, the proof is obvious. So now we assume $t\geq 2$.
We first claim that there exists at least one integer $d$ in $[1,4\delta_1+1]$ such that $\mathcal{C}_{(d,p)}^f< \frac t4$. We prove it by contradiction. Assume for $d=1,\dots,4\delta_1+1$, $\mathcal{C}_{(d,p)}^f\geq \frac t4 $.
Then by Lemma \ref{lm-6}, there exist $\lceil \frac12 \mathcal{C}^f_{(d,p)}\rceil$ polynomials $A_{u,v}(x),u,v\in\{1,\dots,t\},u<v$ such that $A_{u,v}(d)=0$. Since $d=1,\dots,4\delta_1+1$ are different elements in $\F_p$, the polynomials $A_{u,v}(x),u,v\in\{1,\dots,t\},u<v$ have at least $\sum_{d=1}^{4\delta_1+1}\lceil\frac12 \mathcal{C}^f_{(d,p)}\rceil$ roots.
Now $\sum_{d=1}^{4\delta_1+1}\lceil\frac12 \mathcal{C}^f_{(d,p)}\rceil>4\delta_1\lceil\frac12 \cdot\frac t4\rceil\geq\frac12 \delta_1 t=\frac12 (n-1)(T-1)t$, which contradicts to the fact that the sum of the degrees of $\{A_{u,v},u,v\in\{1,2,\dots,t\},u<v\}$ is at most $\frac12 t(t-1)(n-1)$. We proved the claim.
By Lemma \ref{cor-2}, we have $\mathcal{C}_{(d_0,p)}^f\leq 2\mathcal{C}^f_{(d,p)},d=1,\dots,4\delta_1+1$. So $\mathcal{C}_{(d_0,p)}^f<2\cdot\frac t 4=\frac{t}{2}$. So the number of no collision terms of $f$ in $f^{\modp}_{(d_0,p)}$ is at least $>t-\frac t 2=\frac{t}{2}$, which is $\geq \lceil\frac{t}{2}\rceil$. \qed

\subsection{Reduction from multivariate interpolation to univariate interpolation}
Now, we give a reduction algorithm which reduces multivariate interpolation to univariate interpolation, where we assume that a univariate interpolation algorithm exists.
\begin{alg}[MIPolyBase]\label{alg-2}
\end{alg}

{\noindent\bf Input:}
A black-box procedure $\mathcal{B}_f$ that computes  $f\in\RB[\X]$, $T\geq \# f$, $D>\deg(f)$.

{\noindent\bf Output:} The exact form of $f$.

\begin{description}
\item[Step 1:] Let $\delta_1=(n-1)(T-1),\delta_2=(n-1)T,N=\max\{4\delta_1+1,\delta_1+\delta_2+1\}$,
and $p$ a prime such that $p\geq \max\{n,N,D\}$.

\item[Step 2:]
 For $d=1,2,\dots,N$, find
$f_{(d,p)}$ via a univariate interpolation algorithm. Let $f_d=f_{(d,p)},$ $f^{\modp}_d=f^{\modp}_{(d,p)}= f_{(d,p)}\, \modp (x^p-1)$.

\item[Step 3:] Let $\alpha=\max\{\#f^{\modp}_d|d=1,2,\dots,N\}$ and $d_0$ satisfying $\#f^{\modp}_{d_0}=\alpha$. Let $h=0$.

\item[Step 4:] While $\alpha\neq 0$, do

\begin{description}
\item[a:] For $k=1,2,\dots,n$, find $f_{(d_0,p,k)}$ via a univariate interpolation algorithm and let $g_k=f_{(d_0,p,k)}-h_{(d_0,p,k)}$.

\item[b:] Let $\hbox{TS}^{f-h}_{d_0,p,D}\,=\,\mathbf{TSTerms}(f^{\modp}_{d_0},f_{d_0},g_1,\dots,g_n,d_0,p,D)$.

\item[c:] Let $s=0$. For each $u\in \hbox{TS}^f_{d_0,p,D}$,
set $s:=s + u$ if
$$\#\{d\,|\, \#(f^{\modp}_d-u^{\modp}_{(d,p)})<\#f^{\modp}_d,d=1,\dots,\delta_1+\delta_2+1 \}\ge \delta_2+1.$$

\item[d:] Let $h=h+s$, $T=T-\#s$,     $\delta_1=(n-1)(T-1),\delta_2=(n-1)T,$ $N=\max\{4\delta_1+1,\delta_1+\delta_2+1\}$.

\item[e:] For $d=1,\dots,N$, let $f_d=f_d-s_{(d,p)}$, $f^{\modp}_d=f^{\modp}_d-s^{\modp}_{(d,p)}$.

\item[f:] Let $\alpha=\max\{\#f^{\modp}_d|d=1,\dots,N\}$ and $d_0$ satisfying $\#f^{\modp}_{d_0}=\alpha$.

\end{description}

\item[Step 5:] Return $h$

\end{description}

\begin{theorem}\label{the-4}
Algorithm \ref{alg-2} is correct and needs interpolating $O(nT)$ univariate polynomials with degrees less than $O(D\max\{nT,D\})$ and terms less than $T$. Besides this, it still needs $O^\thicksim(nT^2)$ ring operations in $\RB$ and $O^\thicksim(n^2T^2\log D)$ bit operations.
\end{theorem}
\proof
By Theorem \ref{the-2} and Lemma \ref{lm-7},  at least half of terms of $f$ are in $\hbox{TS}^f_{d_0,p}$ obtained in {\bf{b}} of Step 4.
In $\mathbf{c}$ of Step 4, Theorem \ref{the-1} is used to select the elements of $M_f$ from $\hbox{TS}^f_{d_0,p}$.
Then in each loop of Step 4, at least half of the terms of $f$ are obtained and
$f$ will be obtained by running at most $\log_2 T$ loops of Step 4. The correctness of the algorithm is proved.

We now analyse the complexity.
We call $N$ univariate interpolations in Step 2 and at most $n\log_2t$ univariate interpolations in $\mathbf{a}$ of Step 4. Note that $\deg (f_{(d,p)})\leq D(p-1)$ and $\deg f_{(d,p,k)}\leq 2D(p-1)$. Since the $p$ is $\max\{O(nT),O(D)\}$ and $\#f\leq T$, the first part of the theorem is proved.
In Step 2, obtaining one $f_d^{\modp}$ needs $O(t)$ ring operations and $O^\thicksim(t\log (pD))$ bit operations. Since it totally has $N$ polynomials
$f_d$, the complexity is $O(nT^2)$ ring operations and $O^\thicksim(nT^2\log D)$ bit operations.
In Step 3, since $N$ is $O(nT)$ and  $f^{\modp}_{(d_0,p)}$ has no more than $T$ terms, it needs $O^\thicksim(nT^2)$ bit operations.
For Step 4, first we analyse the complexity of one cycle.
In $\mathbf{b}$, by Theorem \ref{the-3}, the complexity is $O(nT)$ ring operations and $O^\thicksim(nT\log D)$ bit operations.
In $\mathbf{c}$, since every check needs to compare $\delta_1+\delta_2+1$ polynomials and to use $\delta_1+\delta_2+1$ substitutions, the complexity is $O(\#\hbox{TS}^f_{d_0,p} n(\delta_1+\delta_2)\log (pD)+\#\hbox{TS}^f_{d_0,p} (\delta_1+\delta_2)\log T\log (pD))$ bit operations and $O(\#\hbox{TS}^f_{d_0,p} (\delta_1+\delta_2))$ ring operations. Since $\#\hbox{TS}^f_{d_0,p}\leq T$, the complexity is $O(nT^2)$ ring operations and $O^\thicksim(n^2T^2\log D)$ bit operations.
In $\mathbf{\mathbf{e}}$, it needs $n(\#s)$ operations to obtain  $s_{(d,p)},s_{(d,p)}^{\modp}$. Since we need to update $N$ polynomials, the complexity is $O^\thicksim(n(\#s)N\log (pD))$ bit operations and $O(\#sN)$ ring operations.
Since  we obtain at least half of the terms in $f$ in every recursion,  at most $\log_2t$ recursions are need.  Since the sum of $\#s$ is $t$, the total complexity of Step 4 is  $O^\thicksim(nT^2)$ ring operations and $O^\thicksim(n^2T^2\log D)$ bit operations. \qed

\subsection{Multivariate interpolation algorithm}
\label{sec-34}
Algorithm \ref{alg-2} reduces multivariate interpolations to univariate
interpolations. In this section, we combine Algorithm \ref{alg-2}
with three univariate interpolation algorithms to give multivariate interpolation algorithms, which are written as corollaries of Theorem \ref{the-4}.

For a univariate polynomial with degree $D$, the Lagrange algorithm works
over any integral domain with more than $D+1$ elements and has arithmetic
complexity $O^\thicksim(D)$ \cite{12}.
Combining Theorem \ref{the-4} with  the Lagrange algorithm, we have the following result.
\begin{cor}\label{cor-lag-1}
Let $\RB$ be an integral domain with more than $O^\thicksim(D\max(nT,D))$ elements.
Then Algorithm \ref{alg-2} needs $O^\thicksim(nTD\max(nT,D))$ queries of $B_f$ and $O^\thicksim(nTD\max(nT,D))$ arithmetic operations over $\RB$, plus a similar number of bit operations if using the Lagrange interpolation algorithm.
\end{cor}

If $\RB$ is the field of complex numbers, we may use the Ben-or and Tiwari algorithm,
whose complexity  is dominated by the root finding step \cite{3,32}. In  \cite[Lemma 2.3 ]{51}, we show that for univariate polynomials, the root can be found by factoring the coefficients of an auxiliary polynomial. As a consequence, the univariate Ben-or and Tiwarri's algorithm costs $O^\thicksim(T\log^2 D)$
$\RB$ operations, which has better complexities than the Lagrange algorithm
since $T \le D$ for univariate polynomials.
%
This result, combining with Theorem \ref{the-4}, gives the following
result.
\begin{cor}\label{cor-b0-1}
Let $\RB = \C$.
Then Algorithm \ref{alg-2} needs $O(nT^2)$ queries of $B_f$  and costs $O^\thicksim(nT^3D$ $\max(nT,D))$ bit operations if  Ben-or and Tiwarri's algorithm is used for univariate interpolation.
\end{cor}
\proof
Since $\#f_{(d,p)},\#f_{(d,p,k)}\leq T$ and $\deg(f_{(d,p)}),\deg(f_{(d,p,k)})$ are $O^\thicksim(D\max(nT,D))$, the arithmetic complexity of interpolating univariate polynomials is $O^\thicksim(nT^2\log^2D)$. Since we evaluate at the points $2^0,\dots,2^{2T-1}$, the height of the data is $O^\thicksim(TD\max(nT,D)$.  So the bit complexity of Algorithm \ref{alg-2} is $O^\thicksim(nT^3D\max(nT,D))$.
\qed

If $\RB=\F_q$, the Ben-or and Tiwari algorithm can be changed into a deterministic algorithm
which needs $2T$ queries and $O^\thicksim(D\log q)$ bit operations \cite{huang-bot}.
Note that the Lagrange algorithm has the same complexity $O^\thicksim(D\log q)$
but needs $O(D)$ queries.
%
%
Based on this modified Ben-or and Tiwari algorithm, we have
\begin{cor}\label{cor-bq-1}
Let $\RB=\F_q$. Then Algorithm \ref{alg-2} needs $O(nT^2)$ queries of $B_f$ and $O^\thicksim(nTD$ $\max(nT,D)\log q)$ extra bit operations if the modified Ben-or and Tiwari's algorithm is used for univariate interpolations.
\end{cor}
\proof
Since the degree of $f_{(d,p)}$ and $f_{(d,p,k)}$ are $O^\thicksim(D\max(nT,D))$, $\#f_{(d,p)},\#f_{(d,p,k)}\leq T$,
by \cite{huang-bot} and Theorem \ref{the-4}, the bit complexity is $O^\thicksim(n^2T^2D\log q)$.\qed

Among the three algorithms, the one in Corollary \ref{cor-lag-1} works
for more general rings, but has high complexities.
Both the algorithms in Corollaries  \ref{cor-b0-1} and  \ref{cor-bq-1}
use Ben-or and Tiwari's algorithm for univariate interpolations,
but work for different coefficient rings.

\section{An interpolation algorithm based on modulus changing Kronecker substitution}
In this section, we give a method to recover $f$ from $f_{(D,p)}$, where $D$ is a degree bound for $f$ and $p$ will be a sufficiently many primes,
that is, we change the modulus $p$ in the Kronecker substitution.
The algorithm is quite similar to that given in Section 3, but we need to
give a different criterion for term testing:
instead of finding a ``ok" degree, we need to find a ``ok" prime.

\subsection{Term testing criterion and ``ok" prime}
We first give a new criterion for term testing.
\begin{lemma}\label{lm-8}
Let $f=\sum_{i=1}^t c_im_i$, $T\geq \# f,D>\deg(f)$, $N_1$ be the smallest number such that $p_1p_2\cdots p_{N_1}\geq D^{n(T-1)}$, where $p_i$ is the $i$-th prime.
Then for each $cm\in M_f$, there exist at most $N_1-1$ primes $p$ such that $cm$ is a collision in $f^{\modp}_{(D,p)}$.
\end{lemma}
\proof
If $t=1$, then $N_1\geq1$. The proof is obvious.
Now we assume $t\geq 2$.
It suffices to show that for any $N_1$ different primes $q_1,q_2,\dots,q_{N_1}$,  there exists at least one $q_j$, such that $cm$ is not a collision in $f^{\modp}_{(D,q_j)}$.
Assume $m_i=x_1^{e_{i,1}}\cdots x_n^{e_{i,n}},i=1,\dots,t$. It suffices to consider the case of $c_1m_1$. We prove it by contradiction.
Assume that for every $q_j,j=1,\dots,N_1$, $c_1m_1$ is a collision in $f^{\modp}_{(D,q_j)}$.
Let
$B=\prod^t_{s=2}(\sum_{i=1}^n(e_{1,i}-e_{s,i})D^{i-1}).$
First, we show that if $c_1m_1$ is a collision in $f^{\modp}_{(D,q_j)}$, then $\modp(B,q_j)=0$.
Since $c_1m_1$ is a collision in $f^{\modp}_{(D,q_j)}$, without loss of generality, assume $m^{\modp}_{1(D,q_j)}=m^{\modp}_{2(D,q_j)}$.
Then $0=\deg (m^{\modp}_{1(D,q_j)})-\deg (m^{\modp}_{2(D,q_j)})=\modp(\sum_{i=1}^ne_{1,i}D^{i-1},q_j)-\modp(\sum_{i=1}^ne_{2,i}\\D^{i-1},q_j)$.
So  $\modp(\sum_{i=1}^n(e_{1,i}-e_{2,i})D^{i-1},q_j)=0$, which implies that $\modp(B,q_j)=0$.
Since $q_1,\dots,q_{N_1}$ are different primes, $\prod^{N_1}_{j=1}q_j$ divides $B$.
Note that $|\sum_{i=1}^n(e_{1,i}-e_{s,i})D^{i-1}|\leq (D-1)(\sum_{i=1}^nD^{i-1})=D^n-1$.
So $|B|=|\prod^t_{s=2}\sum_{i=1}^n(e_{1,i}-s_{s,i})D^{i-1}|\leq (D^n-1)^{t-1}$.
Thus $\prod^{N_1}_{j=1}q_j\geq\prod^{N_1}_{j=1}p_j \geq D^{n(T-1)}>(D^n-1)^{T-1}\geq |B|$, which  contradicts
the fact that $\prod^{N_1}_{j=1}q_j$ divides $B$.
So at least one of $q_j,j=1,\dots,N_1$ does not divide $B$. Without loss of generality, assume $q_1$ does not divide it, then
$\modp(\sum_{i=1}^ne_{1,i}D^{i-1}-\sum_{i=1}^ne_{s,i}D^{i-1},q_1)\neq0$.
So
$\modp(\sum_{i=1}^ne_{1,i}D^{i-1},q_1)\neq \modp\\(\sum_{i=1}^ne_{s,i}D^{i-1},q_1)$, for $s=2,3,\dots,t$. So $c_1m_1$ is not a collision in $f^{\modp}_{(D,q_1)}$.
We proved the lemma.\qed

We give a new criterion for $cm\in C_f$.
\begin{theorem}\label{lm-10}
Let $f=\sum_{i=1}^t c_im_i$, $T\geq \# f,D>\deg f$, $N_1(N_2)$ be the smallest number such that $p_1p_2\cdots p_{N_1}\geq D^{n(T-1)}(p_1p_2\cdots p_{N_2}\geq D^{nT})$, and $p_i$ be the $i$-th prime. For a term $cm$ satisfying $\deg(m)<D$,  $cm\in M_f$ if and only if there exist at least $N_2$ integers $j\in[1,N_1+N_2-1]$ such that $\#(f-cm)^{\modp}_{(D,p_j)}<\#f^{\modp}_{(D,p_j)}$.
\end{theorem}
\proof
Let $cm\in M_f$. If $p_j$ is a prime such that $cm$ is not a collision in $f^{\modp}_{(D,p_j)}$, then $\#(f-cm)^{\modp}_{(D,p_j)}=\#f^{\modp}_{(D,p_j)}-1$. So $\#(f-cm)^{\modp}_{(D,p_j)}<\#f^{\modp}_{(D,p_j)}$.
By Lemma \ref{lm-8}, there exist at most $N_1-1$ primes such that $cm$ is a collision in $f$. In $p_1,\dots,p_{N_1+N_2-1}$, as $N_1+N_2-1-(N_1-1)=N_2$, there exist at least $N_2$ primes such that $\#(f-cm)^{\modp}_{(D,p_j)}<\#f^{\modp}_{(D,p_j)}$.
For the other direction, assume $cm\notin M_f$.
%
Consider two cases:
Case 1: $m$ is not a monomial in $f$.
In this case,  $f-cm$ has at most $T+1$ terms.
By Lemma \ref{lm-8}, there exist at most $N_2-1$ integers $j\in[1,N_1+N_2-1]$ such that $cm$ is a collision in $(f-cm)^{\modp}_{(D,p_j)}$. In $p_1,\dots,p_{N_1+N_2-1}$, since $N_1+N_2-1-(N_2-1)= N_1$, there are at least $N_1$ primes such that $\#(f-cm)^{\modp}_{(D,p_j)}>\#f^{\modp}_{(D,p_j)}$. So there are at most $N_1+N_2-1-N_1=N_2-1$ primes such that $\#(f-cm)^{\modp}_{(D,p_j)}<\#f^{\modp}_{(D,p_j)}$.
Case 2: $m$ is a monomial in $f$, but $cm$ is not a term in $f$. In this case, $f-cm$ has the same number of terms as $f$. Assume the term of $f$ with monomial $m$ is $c_1m$.
By Lemma \ref{lm-8}, there exist at most $N_1-1$ primes such that $(c_1-c)m$ is a collision in $(f-cm)^{\modp}_{(D,p_j)}$. In $p_1,\dots,p_{N_1+N_2-1}$, since $N_1+N_2-1-(N_1-1)=N_2$, there are at least $N_2$ primes such that $\#(f-cm)^{\modp}_{(D,p_j)}=\#f^{\modp}_{(D,p_j)}$. So there are at most $N_1+N_2-1-N_2=N_1-1\leq N_2-1$ primes such that $\#(f-cm)^{\modp}_{(D,p_j)}<\#f^{\modp}_{(D,p_j)}$.\qed

A prime $p$ is called a ``ok" prime, if at least half of terms of $f$ do not collide in $f^{\modp}_{(D,p)}$. We have
%
%
\begin{lemma}\cite{huang-slp}\label{lm-11}
Let $f=\sum_{i=1}^tc_im_i\in\RB[\X]$, $m_i=x_1^{e_{i,1}}\cdots x_n^{e_{i,n}}$, $D>\deg(f)$, $A=\prod_{i,j\in[1,t]}^{i< j}\sum_{k=1}^n$ $(e_{i,k}-e_{j,k})D^{k-1}$,
and $p$ a prime. If $\mathcal{C}^f_{(D,p)}=s$, then $p^{\lceil\frac{s}{2}\rceil}$
divides $A$.
\end{lemma}

The following result gives a method to find an ``ok" prime.
%
\begin{theorem}\label{the-6}
Let $f=\sum_{i=1}^tc_im_i\in\RB[\X]$, $T\geq \# f,D>\deg f$, $N_3$ be the smallest number such that $p_1\cdots p_{N_3}\ge D^{4n(T-1)}$, where $p_i$ is the $i$-th prime. Let
$j_0$ be an integer in $[1,N_3]$ such that $\# f^{\modp}_{(D,p_{j_0})}\geq \# f^{\modp}_{(D,p_{j})}$ for all $j=1,\ldots,N_3$.
%
Then $p_{j_0}$ is an ``ok" prime.
%
\end{theorem}
\proof
We first claim that there exists at least one $p_j$ in $p_1,\dots,p_{N_3}$ such that $\mathcal{C}_{(D,p_j)}^f< \frac t4$. We prove it by contradiction. Assume for $j=1,\dots,N_3$, $\mathcal{C}_{(D,p_j)}^f\geq \frac t4$. Then by Lemma \ref{lm-11}, $p_j^{\lceil\frac12 \mathcal{C}_{(D,p_j)}\rceil}$ divides $A$. Since $p_j,j=1,\dots,N_3$ are different primes, then $\prod_{j=1}^{N_3}p_j^{\lceil\frac12 \mathcal{C}_{(D,p_j)}\rceil}$ divides $A$.

Now $\prod_{j=1}^{N_3}p_j^{\lceil\frac12 \mathcal{C}_{(D,p_j)}\rceil}\geq D^{\frac12 nt(T-1)}$, which contradicts to the fact that $A\leq (D^n-1)^{\frac{t(t-1)}{2}}$. We proved the claim.
By Lemma \ref{cor-2}, we have $\mathcal{C}_{(D,p_{j_0})}^f\leq 2\mathcal{C}^f_{(D,p_{j})},j=1,\dots,N_3$. So $\mathcal{C}_{(D,p_{j_0})}^f<\frac t 2$. So the number of no collision terms of $f$ in $f^{\modp}_{(D,p_{j_0})}$ is $>t-\frac t 2=\frac{t}{2}$, which is at least $\lceil\frac t 2\rceil$.
The theorem is proved.\qed

\subsection{Reduction from multivariate interpolation to univariate interpolation}
We now give the a new reduction algorithm based on the results given in the preceding section.
\begin{alg}[MIPoly2]\label{alg-3}
\end{alg}

{\noindent\bf Input:} A black-box procedure $B_f$ for $f\in\RB[\X]$, $T\geq \# f$, $D>\deg f$.

{\noindent\bf Output:} The exact form of $f$.

\begin{description}
\item[Step 1:] Let $K=\max\{4,\lceil n(T-1)\log_2 D\rceil+\lceil nT\log_2 D\rceil,4\lceil nT\log_2 D\rceil\}$. Let $p_1,\dots,p_{K}$ be the first $K$ primes.

\item[Step 2:]
Find the $N_1,N_2,N_3$ be the integers defined in Lemma \ref{lm-8}, Theorem \ref{lm-10}, Theorem \ref{the-6}. Let $N=\max\{N_1+N_2-1,N_3\}$.%

\item[Step 3:]
 For $j=1,\dots,N$, compute
$f_{(D,p_j)}$ via a univariate interpolation algorithm.
Let $f_j=f_{(D,p_j)},f^{\modp}_j=f^{\modp}_{(D,p_j)}=f_{(D,p_j)}\, \modp (x^p-1)$.

\item[Step 4:] Let $\alpha=\max\{\#f^{\modp}_j|j=1,2,\dots,N\}$ and $j_0$ satisfies $\#f^{\modp}_{j_0}=\alpha$. Let $h=0$.

\item[Step 5:] While $\alpha\neq 0$ do

\begin{description}

\item[a:] For $k=1,2,\dots,n$, compute $f_{(D,p_{j_0},k)}$ via a univariate interpolation algorithm. Let $g_k=f_{(D,p_{j_0},k)}-h_{(D,p_{j_0},k)}$.
%

\item[b:] Let $\hbox{TS}^{f-h}_{D,p_{j_0},D}\!\!\!=\!\!\!\mathbf{TSTerms}(f^{\modp}_{j_0},f_{j_0},g_{1},\dots,g_n,D,p_{j_0},D)$.

\item[c:] Let $s=0$. For each $u\in \hbox{TS}^f_{D,p_{j_0},D}$,
set $s=s + u$ if
$$\#\{j\,|\, \#(f^{\modp}_j-u^{\modp}_{(D,p_j)})<\#(f^{\modp}_j),j=1,\dots,N_1+N_2-1 \}\ge N_2.$$

\item[d:] Let $h=h+s$, $T=T-\#s$, update $N_1,N_2,N_3$ for the new $T$ following Step 2. Let $N=\max\{N_3,N_1+N_2-1\}$.

\item[e:] For $j=1,2,\dots,N$, let $f_j=f_j-s_{(D,p_j)}$, $f^{\modp}_j=f^{\modp}_j-s^{\modp}_{(D,p_j)}$.

\item[f:] Let $\alpha=\max\{\#f^{\modp}_j|j=1,2,\dots,N\}$ and $j_0$ satisfies $\#f^{\modp}_{j_0}=\alpha$.

\end{description}

\item[Step 6:] Return $h$.
\end{description}

\begin{theorem}\label{the-7}
Algorithm \ref{alg-3} is correct and needs interpolating $O(nT\log D)$
univariate polynomials with degrees less than $O^\thicksim(nDT)$ and terms less than $T$. Besides this, we still need $O^\thicksim(nT^2\log D)$ ring operations in $\RB$ and $O^\thicksim(n^2T^2\log D)$ bit operations.
\end{theorem}
\proof
Since $N_1$ is the smallest number such that $p_1\cdots p_{N_1}\geq D^{n(T-1)}$,  we have $2^{N_1-1}\leq p_1 \cdots p_{N_1-1}$ $ <D^{n(T-1)}$. So $N_1\leq \max\{1,\lceil n(T-1)\log_2 D\rceil\}$.
Similarly, we have $N_2\leq \max\{1,\lceil nT\log_2 D\rceil\}$, $N_3\leq 4\{\max(1,\lceil nT\log_2 D\rceil)\}$. So in Step 2, we always have $K\geq N$. By Theorem \ref{the-6} and Lemma \ref{lm-7}, at least half of terms of $f$ are in $\hbox{TS}^f_{D,p_{j_0}}$.
In $\mathbf{c}$ of Step 5, Theorem \ref{lm-10} is used to select the elements of $M_f$ from $\hbox{TS}^f_{D,p_{j_0}}$. The correctness of the algorithm is proved.

We call $N$ univariate interpolations in Step 3 and at most $n\log_2t$ univariate interpolations in  $\mathbf{a}$ of Step 5. Note that $\deg(f_{(D,p_i)})\leq D(p_i-1)$ and $\deg(f_{(D,p_i,k)})\leq2D(p_i-1)$.
Since the $i$-th prime is $O(i\log i)$, $N_1,N_2,N_3$ are $O(nT\log_2 D)$, the first part of the
theorem is proved.

In Step 1, the bit complexity of finding the first $K$ primes is $O^\thicksim(K)$ by \cite[p.500,Them.18.10]{12}. Since $K$ is $O^\thicksim(nT\log D)$, the bit complexity of Step 1 is $O^\thicksim(nT\log D)$.
In Step 2, we need to compute $D^{n(T-1)},D^{nT},D^{4n(T-1)}$. Since the height of data is $O(nT\log D)$ and the arithmetic operation to compute them is $O(\log^2 (nT))$, the bit complexity  is $O^\thicksim(nT\log D)$.
Since $N_1,N_2,N_3$ are $O(nT\log D)$, the bit complexity of finding $N_1,N_2,N_3$ is $O(n^2T^2\log^2 D)$.
%
%
In Step 3, computing one $f_j^{\modp}$ needs $O^\thicksim(T\log (pD))$ bit operations and $O(T)$ ring operations. Since we need to compute $N$ polynomials $f^{\modp}_j$, the complexity is $O^\thicksim(nT^2\log^2 D)$ bit operations and $O(nT^2\log D)$ ring operations.
In Step 4, since $N$ is $O(nT\log D)$ and the terms of $f^{\modp}_{j}$ is no more than $T$, we need $O^\thicksim(nT^2\log D)$ bit operations.
Now we consider Step 5. We first analyse the complexity of one cycle.
In $\mathbf{b}$, by Lemma \ref{the-3}, the complexity is $O^\thicksim(nT\log D)$ bit operations and $O(nT)$ ring operations in $\RB$.
In $\mathbf{c}$, since every check needs to compare $N_1+N_2-1$ polynomials and to use $N_1+N_2-1$ substitutions, the complexity is $O(\#\hbox{TS}^{f-h}_{D,p_{j_0}} n(N_1+N_2)\log (pD)+\#\hbox{TS}^{f-h}_{D,p_{j_0}} (N_1+N_2)\log T\log(pD))$ bit operations and $O(\#\hbox{TS}^{f-h}_{D,p_{j_0}} (N_1+N_2))$ ring operations. Since $\#\hbox{TS}^{f-h}_{D,p_{j_0}}\leq T$, the complexity is $O^\thicksim(n^2T^2\log^2 D)$ bit operations and $O(nT^2\log D)$ ring operations.
In $\mathbf{d}$, for the same reason as in Step 2, the complexity is $O(n^2T^2\log^2D)$ bit operations.
In $\mathbf{e}$,  we need $n\# s$ operations to obtain $s_{(D,p_j)}$ and $s_{(D,p_j)}^{\modp}$.
 Since we need to update $N$ polynomials, the complexity is $O^\thicksim(n(\#s)N\log (pD))$ bit operations and $O(\#sN)$ ring operations.
Since in every recursion, we obtain at least half of the terms in $f$,  at most $\log_2t$ recursions are needed.  Since the sum of $\#s$ is $t$, the complexity of Step 5 is  $O^\thicksim(n^2T^2\log^2 D)$ bit operations and $O^\thicksim(nT^2\log D)$ ring operations. \qed

\subsection{Multivariate interpolation}
Similar to section \ref{sec-34}, we gave three multivariate interpolation
algorithms by combining Algorithm \ref{alg-3} with three different univariate interpolation algorithms.

Similar to Corollary \ref{cor-lag-1}, we have the following result.
\begin{cor}\label{cor-lag-2}
Let $\RB$ be an integral domain with more than $O^\thicksim(nTD)$ elements
and $f\in\RB[\X]$.
Then Algorithm \ref{alg-3} needs $O^\thicksim(n^2T^2D)$ queries of $B_f$ and $O^\thicksim(n^2T^2D)$
arithmetic operations over $\RB$ plus a similar number of bit operations if using the Lagrange interpolation algorithm.
\end{cor}

Similar to Corollary \ref{cor-b0-1}, we have the following result.
\begin{cor}\label{cor-b0-2}
Let $\RB=\C$. Then
Algorithm \ref{alg-3} needs $O(nT^2\log D)$ queries of $B_f$ and $O^\thicksim(n^2T^4D)$
 bit operations if Ben-or and Tiwarri's algorithm is used for univariate interpolations.
\end{cor}

Similar to  Corollary \ref{cor-bq-1}, we have the following result.
\begin{cor}\label{cor-bq-2}
Let $\RB=\F_q$. Then
Algorithm \ref{alg-3} needs $O(nT^2\log D)$ queries of $B_f$ and $O^\thicksim(n^2T^2$ $D\log q)$ bit operations  if  Ben-or and Tiwari's algorithm is used for univariate interpolations.
\end{cor}

\section{Conclusion}
In this paper, we revisit the approach of reducing
the black-box multivariate polynomial interpolation to that of
the univariate polynomial by introducing a new
modified Kronecker Substitution.
Over the field $\C$, the bit complexity of the algorithm
is linear in $D$, while all existing deterministic algorithms are quadratic in $D$.
Over finite fields, the new algorithm has better complexities comparing to existing deterministic algorithms in $T$ and $D$ and has the same complexity in $n$.

Finally, we compare the complexities of Algorithm \ref{alg-2} and Algorithm \ref{alg-3}
in the case of finite fields. In other cases, the results are the same.
The complexities of Algorithm \ref{alg-2} and Algorithm \ref{alg-3}
are $O^\thicksim(nTD$ $\max(nT,D)\log q)$ and $O^\thicksim(n^2T^2$ $D\log q)$,
respectively.
If $nT>D$, the two algorithms have the same complexities.
If $nT<D$, the ratio of the complexities of Algorithm \ref{alg-2} and Algorithm \ref{alg-3}
is $\frac{D}{nT}$, that is, Algorithm \ref{alg-3} performs better for large $D$.


\begin{thebibliography}{99}
\bibitem{AM95}
N. Alon, Y. Mansour,
\newblock Epsilon-discrepancy sets and their application for interpolation of sparse polynomials,
\newblock Inform. Process. Lett. 54(6) (1995) 337-342.

\bibitem{1}
A. Arnold and D.S. Roche
\newblock ``Multivariate sparse interpolation using randomized Kronecker substitutions,"
\newblock ISSAC'14, ACM Press, 35-42, 2014.

\bibitem{2}
A. Arnold, M. Giesbrecht, D.S. Roche,
\newblock ``Faster sparse multivariate polynomial interpolation of straight-line programs,"
\newblock J. of Sym. Comp., 75, 4-24, 2016.

\bibitem{21}
A. Arnold, M. Giesbrecht, D.S. Roche,
\newblock  ``Faster sparse interpolation of straight-line programs,"
\newblock In CASC13, LNCS 8136, 61-74. 2013



\bibitem{s10}
M. Avenda\~{n}o, T. Krick,  A. Pacetti,
\newblock ``Newton-Hensel interpolation lifting,"
\newblock Foundations of Computational Mathematics, 6(1), 82-120, 2006.


\bibitem{3}
M. Ben-Or and P. Tiwari,
\newblock ``A deterministic algorithm for sparse multivariate polynomial interpolation,"
\newblock Proc. STOC'88, ACM Press, 301-309, 1988.

\bibitem{621}
M. Clausen, A. Dress, J. Grabmeier, M. Karpinski,
\newblock ``On zero-testing and interpolation of $k$-sparse multivariate polynomials over finite fields,"
\newblock Theoretical Computer Science, 84, 151-164, 1991.

\bibitem{lee-new}
A. Cuyt,  and W.S. Lee,
\newblock ``A new algorithm for sparse interpolation of multivariate polynomials,"
\newblock Theoretical Computer Science, 409(2), 180-185, 2008.


\bibitem{5}
S. Garg and E. Schost,
\newblock ``Interpolation of polynomials given by straight-line programs,"
\newblock Theoretical Computer Science, 410, 2659-2662, 2009.

\bibitem{8}
M. Giesbrecht and D.S. Roche,
\newblock ``Diversification improves interpolation,"
\newblock  Proc. ISSAC'11,  ACM Press, 123-130, 2011.

\bibitem{GR10}
M.Giesbrecht, D.S. Roche,
\newblock ``Interpolation of Shifted-Lacunary Polynomials[J],"
\newblock Computational Complexity, 2010, 19(3):333-354.

\bibitem{BJ14}
M.Bl\"{a}ser, G.Jindal,
\newblock ``A new deterministic algorithm for sparse multivariate polynomial interpolation[C]"
\newblock Proceedings of the 39th International Symposium on Symbolic and Algebraic Computation. ACM, 2014: 51-58.

\bibitem{14}
M. Giesbrecht, G. Labahn, W. Lee,
\newblock Symbolic-numeric sparse interpolation of multivariate polynomials.
\newblock Proc. ISSAC'06, ACM Press, 116-123, 2006 .

\bibitem{101}
D.Y. Grigoriev, M. Karpinski, M.F. Singer,
 \newblock ``Fast parallel algorithms for sparse multivariate polynomial interpolation over finite fields,"
 \newblock SIAM J. on Comput., 19, 1059-1063, 1990.

\bibitem{16}
M.D.A. Huang and A.J. Rao,
\newblock ``Interpolation of sparse multivariate polynomials over large finite fields with applications,"
\newblock Journal of Algorithms, 33, 204-228, 1999

\bibitem{51}
Q.L. Huang,
 \newblock ``An improved early termination sparse interpolation algorithm for multivariate polynomials,"
 \newblock J. of Sys. Sci. $\&$ Comp., 16, 1-13, 2017.


\bibitem{huang-slp}
Q.L. Huang and X.S. Gao,
\newblock ``Asymptotically optimal Monte Carlo sparse multivariate polynomial interpolation algorithms of straight-line program,"
\newblock arXiv 1709.08979v2, 2017.

\bibitem{huang-bot}
Q.L. Huang and X.S. Gao,
\newblock ``Revisit Randomized Kronecker Substitution based Sparse Polynomial Interpolation,"
\newblock arXiv 1712.05481, 2017.




\bibitem{17}
 S.M.M. Javadi and M. Monagan,
 \newblock ``Parallel sparse polynomial interpolation over finite fields,"
 \newblock Proc. PASCO '10, ACM Press, 160-168, 2010.


\bibitem{32}
E.L. Kaltofen and Y.N. Lakshman,
\newblock ``Improved sparse multivariate polynomial interpolation algorithms,"
\newblock Proc. ISSAC'88, Springer-Verlag, 467-474, 1988.

\bibitem{kal-m}
 E.L. Kaltofen, Y.N. Lakshman, J.M. Wiley,
 \newblock ``Modular rational sparse multivariate polynomial interpolation,"
 \newblock Proc. ISSAC'90, ACM Press, 135-139, 1990.

\bibitem{kal-e1}
E.L. Kaltofen and W.S. Lee,
\newblock ``Early termination in sparse interpolation algorithms,"
\newblock Journal of Symbolic Computation, 36, 365-400, 2003.

\bibitem{15}
 E.L. Kaltofen, W.S. Lee, Z. Yang,
 \newblock ``Fast estimates of hankel matrix condition numbers and numeric sparse interpolation,"
 \newblock  SNC '11, ACM Press, 130-136, 2011.

\bibitem{9}
A.R. Klivans and D. Spielman,
\newblock `` Randomness efficient identity testing of multivariate polynomials,"
\newblock  Proc. STOC '01,  ACM Press, 216-223, 2001.


\bibitem{11}
Y.N. Lakshman and B.D. Saunders,
\newblock ``Sparse polynomial interpolation in nonstandard bases,"
\newblock  SIAM J. Comput., 24(2), 387-397, 1995.

\bibitem{M95}
Y. Mansour,
\newblock Randomized interpolation and approximation of sparse polynomials,
\newblock SIAM J. Comput. 24 (2) (1995) 357-368.


\bibitem{12}
J. von zur Gathen and J. Gerhard,
\newblock ``Modern Computer Algebra,"
\newblock Cambridge University Press, 1999.

\bibitem{13}
R. Zippel,
\newblock ``Interpolating polynomials from their values,"
\newblock J. of Symb. Comp., 9, 375-403, 1990.

\end{thebibliography}
\end{document}